\documentclass[aps,10pt,twocolumn,showpacs,pra,citeautoscript,amsmath,amssymb,floatfix,superscriptaddress]{revtex4-1}

\usepackage{amsmath}
\usepackage{amssymb}
\usepackage{epsfig}
\usepackage{color,epstopdf}
\usepackage{amsthm, thm-restate}
\usepackage{hyperref}

\usepackage{soul}
\usepackage[caption=false]{subfig}
\usepackage{algpseudocode}

\newtheorem{theorem}{Theorem}
\newtheorem*{theorem*}{Theorem}

\newtheorem*{lemma*}{Lemma}

\newtheorem{definition}{Definition}
\newtheorem{conjecture}{Conjecture}

\newcommand{\ket}[1]{|#1\rangle}

\newcommand{\ketbra}[2]{|#1\rangle\langle #2|}
\newcommand{\tr}[2]{\mathrm{Tr}_{#1}\left[ #2 \right]}

\begin{document}
\title{Coherence and asymmetry cannot be broadcast}

\author{Matteo Lostaglio}
\affiliation{ICFO-Institut de Ciencies Fotoniques, The Barcelona Institute of Science and Technology, Castelldefels (Barcelona), 08860, Spain}
\author{Markus P.\ M\"uller}
\email{markusm23@univie.ac.at}
\affiliation{Institute for Quantum Optics and Quantum Information,
Austrian Academy of Sciences, Boltzmanngasse 3, A-1090 Vienna, Austria}
\affiliation{Perimeter Institute for Theoretical Physics, 31 Caroline Street North, Waterloo, Ontario N2L 2Y5, Canada}

\date{April 4, 2019}

\begin{abstract}
In the presence of conservation laws, superpositions of eigenstates of the corresponding conserved quantities cannot be generated by quantum dynamics. Thus, any such coherence represents a potentially valuable resource of asymmetry, which can be used, for example, to enhance the precision of quantum metrology or to enable state transitions in quantum thermodynamics.
Here we ask if such superpositions, already present in a reference system, can be broadcast to other systems, thereby distributing asymmetry indefinitely at the expense of creating correlations. We prove a no-go theorem showing that this is forbidden by quantum mechanics in every finite-dimensional system. In doing so we also answer some open questions in the quantum information literature concerning the sharing of timing information of a clock and the possibility of catalysis in quantum thermodynamics. We also prove that even weaker forms of broadcasting, of which \AA berg's `catalytic coherence' is a particular example, can only occur in the presence of infinite-dimensional reference systems. Our results set fundamental limits to the creation and manipulation of quantum coherence and shed light on the possibilities and limitations of quantum reference frames to act catalytically without being degraded.
\end{abstract}

\maketitle

\emph{Introduction.} Cloning and broadcasting, with their associated no-go theorems, are central results marking the difference between classical and quantum information. No cloning is the result that there is no machine that takes as input an unknown member of a family of distinct quantum states \mbox{$\mathcal{S}=\{\rho^i_S\}_{i=1}^n$} and outputs two independent copies of it: $\rho^i_S \mapsto \rho^i_{S} \otimes \rho^i_{S'}$ for every $i$ (unless $n=1$ or the states are mutually orthogonal) \cite{scarani2005quantum}. Broadcasting is a generalization of this task, in which we require the machine to only `locally clone' the state: $\rho^i_{S} \mapsto \sigma^i_{SS'}$ with $\tr{S}{\sigma^i_{SS'}} = \rho^i_{S'}$, $\tr{S'}{\sigma^i_{SS'}} = \rho^i_{S}$, for every $i$. The no-broadcasting theorem says that this is possible if and only if the states in $\mathcal{S}$ are mutually commuting \cite{barnum1996noncommuting}.

The ability to create quantum states that are in a superposition of eigenstates of an observable is a major paradigm shift distinguishing classical and quantum theories. An archetypical example is the generation of states that are in a coherent superposition of different eigenstates of the Hamiltonian $H_S$. These can be used as resources for metrology \cite{marvian2016how} and quantum thermodynamics~\cite{lostaglio2015description}, and they determine quantum speed limits \cite{marvian2016quantum}. In this work we extend the notions of cloning and broadcasting to ask: can a superposition (or `quantum coherence') be cloned or broadcast? More generally, can other forms of asymmetry with respect to a group representation be broadcast? Here we show that this is forbidden by the laws of quantum mechanics, thereby solving certain open problems in the quantum information and thermodynamics literature \cite{JanzingBeth, Malabarba, Mueller2017, faist2018fundamental, cirstoiu2017irreversibility}. Connecting to recent results~\cite{aberg2014catalytic}, we also show that even weaker forms of broadcasting are allowed by quantum theory only in the presence of infinite-dimensional reference systems. Our no-go theorems apply to the task of broadcasting a \emph{single known} superposition when the dynamics is \emph{restricted} by a conservation law; it hence complements the no-go theorem derived for the creation of superpositions of \emph{multiple unknown} states through \emph{unrestricted} dynamics \cite{oszmaniec2016creating}. 
  
  \begin{figure}
  	\centering \includegraphics[width=\columnwidth]{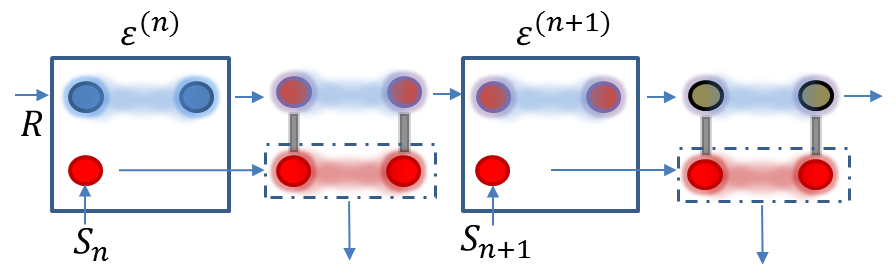}
  	\caption{\emph{A weak broadcasting protocol} (see Def.~\ref{def:weakbroad}). A reference system $R$ distributes asymmetry/coherence to a system $S_n$, transforming it from an eigenstate to a superposition of eigenstates. The state of $R$ gets correlated to $S_n$ and is allowed to change, but it must be able to induce the same transition on a fresh copy of the system $S_{n+1}$ for all $n$. \r{A}berg's protocol~\cite{aberg2014catalytic} is a special case of weak broadcasting of coherence with \mbox{$\dim R=\infty$}.}
  	\label{fig:weak}
  \end{figure}

\emph{Quantum coherence and conservation laws.}  In the original no-go theorems, the constraint that makes cloning and broadcasting nontrivial is that we ask for a single machine to accomplish the task for every state within $\mathcal{S}$. In the case of cloning/broadcasting a single superposition, which is the notion we want to formalize here, what makes the problem nontrivial is the presence of conservation laws. Hence, let us discuss conservation laws in more detail.

Suppose that we have a closed system whose unitary dynamics $U$ is restricted by conservation laws $[U,O_i]=0$, where the $O_i$ are conserved quantities. In this case, $U$ cannot create coherent superpositions of eigenstates of the $O_i$ from single eigenstates, or from incoherent mixtures of such eigenstates.
 This state of affairs admits a more general description in terms of symmetries.  
We have a connected Lie group $G$, and, for any $g\in G$, a unitary $U_g$ such that $g\mapsto U_g$ defines a continuous (possibly projective) group representation.
The conservation laws can be seen as arising from the symmetry of the unitary dynamics $U$, which must satisfy
\begin{equation}
\label{eq:symmetricU}
 [U, U_g] = 0 \quad \textrm{ for every } \; g \in G.
\end{equation} 
The states $\rho_S$ that are a coherent superposition of the corresponding charges (energy, angular momentum, etc.) are those that satisfy $\mathcal{U}_g(\rho_S):= U_g \rho_S U_g^\dag \neq \rho_S$ for at least one $g\in G$. Since these are not invariant under the action of the symmetry, they can be called \emph{G-asymmetric}, otherwise they are called \emph{G-symmetric}~\cite{bartlett2007reference}. A $G$-symmetric state cannot be made asymmetric if the unitary dynamics is itself symmetric with respect to $G$, i.e.
\begin{equation}
\label{eq:impossibility}
\rho_S(g):=\mathcal{U}_g(\rho_S) = \rho_S \Rightarrow  \mathcal{U}_g(U\rho_SU^\dag ) = U\rho_SU^\dagger
\end{equation} 
if $U$ satisfies Eq.~\eqref{eq:symmetricU}.

The impossibility of creating superpositions of different energy levels when the dynamics preserves energy is a particular instance of the the impossibility of creating asymmetric states from symmetric dynamics. To see this explicitly, note that in the case of the group of time translations generated by $H_S$, the states that have nonzero off-diagonal terms in the energy eigenbasis are exactly those $\rho_S$ not invariant under time translations, $\mathcal{U}_t(\rho_S) \neq \rho_S$ for some $t$, where $U_t = e^{-i H_S t}$ (equivalently, \mbox{$[\rho_S,H_S] \neq 0$}). Formally, this can be seen as a representation of the Abelian group $G=\mathbb{R}$ via time translations. This definition of coherence must be distinguished from different notions employed in other works, where coherence sometimes refers to superpositions in an arbitrary basis (see the Scope section in the Supplemental Material).

\emph{Extension to open quantum systems.} The above characterization is easily extended to open quantum system dynamics \cite{bartlett2007reference, marvian2014extending}. In fact, one can introduce an ancilla $A$ with no quantum asymmetry and assume a conservation law holds on $SA$, i.e.,
\begin{equation}
\label{eq:Gcovariant}
\mathcal{E}(\rho_S) := \tr{A}{U(\rho_S \otimes \rho_A)U^\dag }, 
\end{equation}
with $[U,U_g \otimes U^{A}_g] = 0$ $\forall g \in G$ if $U^A_g$ is the action of the symmetry group $G$ on the ancilla, and $\rho_A$ satisfies $\mathcal{U}^A_g(\rho_A) :=U^A_g\rho_A U^A_g = \rho_A$ $\forall g \in G$. The channels defined in Eq.~\eqref{eq:Gcovariant} are called \emph{G-covariant} (or \emph{symmetric}) and they can equivalently be characterised as those satisfying (\cite[Theorem 25]{marvianthesis}, \cite{KeylWerner})
\[
[\mathcal{E}, \mathcal{U}_g] = 0  \quad \textrm{ for every } \; g \in G.
\]
While this is formally analogous to Eq.~\eqref{eq:symmetricU}, note that there is in general no conservation law on $S$ of the form of Eq.~\eqref{eq:symmetricU}; it is rather the global unitary $U$ acting on both system and environment that satisfies a conservation law. Returning to the case where $G=\mathbb{R}$ is the group of time translations, if the dynamics is $\mathbb{R}$-covariant (better known as phase covariant), then 1.\ the ancilla has Hamiltonian $H_A$ and is in a state $\rho_A$ with $[\rho_A,H_A ] =0$ (i.e., incoherent in the energy basis) and 2.\ the global unitary $U$ satisfies overall energy conservation, $[U,H_S + H_A] = 0$. An example of $\mathbb{R}$-covariant maps are thermal operations, defining the resource theory approach to thermodynamics~\cite{janzing2000thermodynamic, HorodeckiOppenheim,Brandao,LostaglioNotes}:
\begin{equation}
\label{eq:thermalops}
\mathcal{E}_T(\rho_S) := \tr{A}{U\left(\rho_S \otimes \gamma_A \right)U^\dag }, 
\end{equation}
where $ \gamma_A = e^{-\beta H_A}/{\tr{}{e^{-\beta H_A} }}$ and $\beta\geq 0$ is some fixed inverse background temperature.

Covariant open dynamics generalize symmetric unitaries and still cannot create coherence:
\begin{equation}
\label{eq:impossibilityopen}
\mathcal{U}_g(\rho_S) = \rho_S \; \; \forall g \in G \Rightarrow  \mathcal{U}_g(\mathcal{E}( \rho_S)) = \mathcal{E}(\rho_S) \; \; \forall g \in G
\end{equation} 
if $\mathcal{E}$ is a symmetric channel. This generalizes Eq.~(\ref{eq:impossibility}).

Furthermore, if in Eq.~(\ref{eq:Gcovariant}) we allowed asymmetric unitaries or asymmetric ancillas, then it would be simple to create an unlimited amount of coherence. For the example of time translations (say, with $H_S$ equal to the Pauli $Z$ matrix for concreteness), a trivial example of the first kind is to perform a Hadamard unitary on $S$ (which does not conserve energy); and a trivial example of the second kind is to introduce an ancilla with Hamiltonian $H_A = H_S$ and in a state $(\ket{0} + \ket{1})/\sqrt{2}$, and then to perform a swap on $SA$.     

 \emph{Coherence broadcasting.} To possess a system $R$ in a state displaying quantum coherence --- or more generally breaking a symmetry --- is a resource that allows us to lift the constraints imposed by conservation laws (sometimes $R$ is called a \emph{reference frame} and the constraint a \emph{superselection rule}) \cite{bartlett2007reference, marvian2014extending}. Various studies noted that the use of quantum coherence as a resource, for example to generate coherence in other subsystems, seems to always come with a degradation of the coherence source \cite{bartlett2007reference, ahmadithesis, BartlettJModOpt, BartlettNJP}. This intuitive picture has partially been put into question by a recent result of \AA berg, who proved the existence of a protocol exhibiting a form of `coherence catalysis' \cite{aberg2014catalytic}, in which a `coherent seed' can be used to generate coherence in arbitrarily many subsystems that get correlated in the process. This protocol can be applied to the task of work extraction from coherence \cite{aberg2014catalytic,korzekwa2016extraction}, and it led to further theoretical insights on the distribution of coherence and asymmetry \cite{cirstoiu2017irreversibility}. However, \AA berg's construction uses an infinite-dimensional reference system whose support spreads in energy at each use without bound, which also requires control of an arbitrarily large number of energy levels. This raises the question of whether the same functionality can be attained with a finite reference.
  
 Here we clarify the situation by proving three general no-go theorems. Our results illuminate how coherence or asymmetry can or cannot be distributed in a catalytic fashion. Given the above discussion on conservation laws, we define the notions of cloning and broadcasting of asymmetry. We start with the thermodynamically relevant case of energetic coherence, then extend to general symmetry groups and subsequently consider weaker forms of broadcasting as illustrated in Fig.~\ref{fig:weak}. To state the definition, we use the notation $\rho_R|\rho'_S$ to denote an arbitrary bipartite state $\sigma_{RS}$ (in general correlated or entangled) that has reduced states $\rho_R={\rm Tr}_S\sigma_{RS}$ and $\rho'_S={\rm Tr}_R \sigma_{RS}$.
\begin{definition}[Coherence cloning and broadcasting]
\label{DefCoh}
We say that the coherence of $\rho_R$ can be \emph{broadcast} if there exists a state $\rho_S$ with $[\rho_S,H_S]=0$ and a time-translation covariant channel $\mathcal{E}$ with
\[
   \mathcal{E}(\rho_R\otimes\rho_S)=\rho_R |\rho'_S,
\]
such that $[\rho'_S,H_S]\neq 0$. We say that the coherence in $\rho_R$ is \emph{cloned} if, in addition, the output is uncorrelated, i.e.
\begin{equation}
		\label{eq:strictcatalysis}
	\mathcal{E} (\rho_R \otimes \rho_S ) = \rho_R \otimes \rho'_S.
\end{equation}
\end{definition}
In all of this work, we restrict ourselves to the case that both $R$ and $S$ are finite-dimensional, unless specified otherwise.

Some comments are in order. First, note that we define coherence broadcasting in a very general way: we only require $\rho'_S$ to have \emph{some} coherence, rather than having in any sense `the same' coherence as $\rho_R$. Cloning of coherence can also be understood as a form of \emph{catalysis} in the usual sense of resource theories, in which $R$ is required to be unchanged and uncorrelated with $S$~\cite{jonathan1999entanglement}. Second, note that coherence cloning implies coherence broadcasting, but the converse is not true: broadcasting allows correlations to be created between $S$ and $R$. Third, coherence broadcasting is in a precise sense a generalization of the notion of broadcasting of the states $\{\rho_R(t)\}_{t \in \mathbb{R}}$, as we show in the Supplemental Material. It is unrelated, however, to the notions used in Refs.~\cite{chakrabarty, bu2016catalytic}.

Using a construction by Janzing et al.~\cite{JanzingBeth}, it is easy to see that coherence cloning is impossible (see also~\cite{vaccaro2018coherence}). For a quantum system with Hamiltonian $H$ in state $\rho$, define the \emph{quantum Fisher information} as $I(\rho,H):={\rm tr}(\dot\rho \Delta_{\rho}^{-1}\dot\rho)$, where $\Delta_\rho B:=(\rho B+B\rho)/2$ and $\dot \rho:=i[\rho,H]$. Then $I$ is non-increasing under covariant maps $\mathcal{E}$, i.e.\ $I(\mathcal{E}(\rho),H)\leq I(\rho,H)$, and it is additive on tensor products, i.e.\ $I(\rho\otimes\tilde\rho,H\otimes\mathbf{1}+\mathbf{1}\otimes \tilde H)=I(\rho,H)+I(\tilde \rho,\tilde H)$. Eq.~(\ref{eq:strictcatalysis}) implies that $I(\rho_R,H_R)+I(\rho'_S,H_S)=I(\rho_R\otimes\rho'_S,H_{RS})\leq I(\rho_R\otimes\rho_S,H_{RS})=I(\rho_R,H_R)+I(\rho_S,H_S)=I(\rho_R,H_R)$. Hence $I(\rho'_S,H_S)=0$, and so $[\rho'_S,H_S]=0$, contradicting Definition~\ref{DefCoh}.

Yet, impossibility of broadcasting cannot be proven as easily: there are correlated states $\rho_{RS}$ such that $I(\rho_R,H_R)+I(\rho_S,H_S)>I(\rho_{RS},H_{RS})$, i.e.\ the sum of the local coherences, as measured by the Fisher information, can be \emph{larger} than the global coherence \cite{hansen2007wigner, cai2008weak}. In other words, correlations can decrease the total amount of coherence, which is what makes broadcasting seem less unlikely than cloning. While $I$ can be used to rule out stronger versions of coherence distribution, the techniques of~\cite{JanzingBeth} are not sufficient to rule out coherence broadcasting in the sense of our definition.

Nevertheless, here we show that

\begin{theorem}
\label{TheCoherence}
Quantum coherence can neither be cloned nor broadcast.
\end{theorem}
This theorem will follow as a special case (for $G=\mathbb{R}$) from Theorem~\ref{thm:noasymmetrybroadcasting} below. For completeness, we give a self-contained proof in the Supplemental Material.

\emph{Application: no broadcasting of timing information beyond classical limit.} One way to interpret the result is that there is no way of using a state $\rho_R$ that is sensitive to time translations (a `clock') to distribute timing information into another system $S$, without affecting the state of the clock. This holds even if correlations are allowed to build up among system and clock. This can be contrasted with the classical limit, represented by the limiting case in which $\rho_R$ is an \emph{unbounded} reference \cite{bartlett2007reference} (one for which the $\{\rho_R(g)\}_{g \in G}$ are mutually orthogonal). This is exemplified by a coherent state $\ket{\alpha} \propto \sum_{n=0}^{\infty} \alpha^n/\sqrt{n!} \ket{n}$, which in the limit $|\alpha| \rightarrow \infty$ becomes an arbitrarily good clock. Since the states $\{\ket{\alpha(t)}\}_{t \in \mathbb{R}}$ become arbitrarily close to mutually orthogonal, the timing information can be cloned arbitrarily well in the limit (see Supplemental Material). In other words, classical clocks (idealized infinite limits of quantum clocks) can distribute timing information without being degraded, but no finite-dimensional quantum clock can. We will see in Theorem~\ref{thm:infinite} that even weaker forms of distribution of time information require infinite clocks. 

Is coherence broadcasting as defined above possible in  infinite-dimensional systems? We leave this question open, as the answer may depend on the details of the mathematical formulation of the problem (see additional comments in the Supplemental Material).

\emph{Application: coherent limitations of correlating thermal machines.} Our results also resolve an open problem from Ref.~\cite{Mueller2017}. There, it was shown that for any given pair of \emph{block-diagonal} states $\rho_A,\rho'_A$, there exists another system $R$ (a `machine') in state $\sigma_R$ such that
\[
\rho_A\otimes\sigma_R\mapsto \rho'_A(\epsilon)|\sigma_R
\]
via a thermal operation, where $\rho'_A(\epsilon)$ is arbitrarily close to the desired target state $\rho'_A$, if and only if $F(\rho_A)\geq F(\rho'_A)$. Here $F$ is the non-equilibrium quantum free energy, $F(\sigma_X) = \tr{}{\sigma_X H_X}  - k_B T S(\sigma_X)$, with $k_B$ Boltzmann's constant, $S$ the von Neumann entropy, and $T$ the ambient temperature. This is in stark contrast to the `infinite second laws' that need to hold when no correlations between $A$ and $R$ are allowed at the output \cite{Brandao}. In that paper, it was conjectured that this statement remains true also in the fully quantum regime, i.e.\ if $\rho'_A$ is not block-diagonal (but $\rho_A$ possibly is). However, Theorem~\ref{TheCoherence} disproves this conjecture: if $\rho_A$ is incoherent then so is $\rho'_A$. In other words, correlating thermal machines are quantum-limited. 

\emph{Application: no thermodynamic generation of coherence without disturbance.}
Recent results have established the fundamental work cost of quantum processes under the assumption that Gibbs-preserving maps (i.e.\ the set of all channels satisfying $\mathcal{E}_G (e^{-\beta H_S}/\tr{}{e^{-\beta H_S}})=e^{-\beta H_S}/\tr{}{e^{-\beta H_S}}$ for some fixed $\beta \geq 0$) can be performed at no work cost. However, the question was raised of what their coherent cost is~\cite{faist2018fundamental}. Suppose that we achieve $\mathcal{E}_G$ by an energy preserving unitary $U$ as
\[
\mathcal{E}_G (\rho_S) = \tr{RB}{U (\rho_R \otimes \ketbra{W}{W}_B \otimes \rho_S)U^\dagger}, 
\]
using a battery $B$ with some Hamiltonian $H_B$ such that $H_B|W\rangle = W |W\rangle$, a coherence source $R$ with Hamiltonian $H_R$ and state $\rho_R$, while \mbox{$[U,H_R+H_B+H_S]=0$}. Then assume $[\rho_S, H_S] = 0$ and \mbox{$[\mathcal{E}_G(\rho_S),H_S] \neq 0$}, i.e. $\mathcal{E}_G$ creates coherence (this is possible \cite{faist2015gibbs}). Since $\mathcal{E}(\rho_R \otimes \rho_S) := \tr{B}{U (\rho_R \otimes \ketbra{W}{W}_B \otimes \rho_S)U^\dagger}$ is $\mathbb{R}$-covariant, it follows from the no coherence broadcasting theorem that the state in $R$ must change,
regardless of whether, or how much, energy from the battery $B$ is consumed in the process (the same holds if, instead of $U$, we use a generic thermal operation).

\emph{Asymmetry broadcasting.} The above results are not restricted to coherence in the basis of a single observable like $H_S$. They generalize as follows:
\begin{definition}[Asymmetry broadcasting]
\label{DefAsymmetry}
 We say that $R$ can broadcast $G$-asymmetry if there exists a system $S$ in a $G$-symmetric state $\rho_S$ and a $G$-covariant operation $\mathcal{E}$ on $RS$ such that
\[
   \mathcal{E}(\rho_R\otimes\rho_S)=\rho_R|\rho'_S
\]
	with $\rho'_S$ a $G$-asymmetric state.
\end{definition}

\begin{theorem}\label{thm:noasymmetrybroadcasting}
$G$-asymmetry broadcasting is impossible for every connected Lie group $G$.
\end{theorem}
For example, rotationally symmetric dynamics cannot be used to broadcast coherence among angular momentum states from an initial superposition of such states. In the language of quantum reference frames \cite{bartlett2007reference}, the above result says that the asymmetry of a quantum reference frame $\rho_R$ cannot be broadcast. Of course, this also implies the simpler result that a quantum reference frame cannot be cloned.

A detailed proof is given in the Supplemental Material, but the main idea is as follows. If $\mathcal{E}$ is $G$-covariant as in Definition~\ref{DefAsymmetry}, then $[\mathcal{U}_g^R\otimes\mathcal{U}_g^S,\mathcal{E}]=0$, hence $\mathcal{E}(\rho_R(g)\otimes\rho_S)=\mathcal{E}\circ \mathcal{U}_g^R\otimes\mathcal{U}_g^S(\rho_R\otimes\rho_S)=\mathcal{U}_g^R\otimes \mathcal{U}_g^S(\rho_R |\rho'_S)=\rho_R(g)|\rho'_S(g)$. While this is not literally broadcasting ($\rho'_S(g)$ is not in general a copy of $\rho_R(g)$), this shows that $\mathcal{E}$ distributes some quantum information of $R$ into $S$: the $\rho_R(g)$ are not all perfectly distinguishable, yet some information that potentially helps distinguish them is transferred to $S$, while the $\rho_R(g)$ are exactly preserved. Intuitively, this seems impossible, and this intuition can be made rigorous by employing what is known as the Koashi-Imoto decomposition~\cite{KoashiImoto,Hayden2004}; see the Supplemental Material.

Similarly, connectedness of the Lie group $G$ is crucial.  $G$-asymmetry for discrete groups $G$ can be broadcast in some cases. For example, consider the case $G=S_n$, the permutation group on $n$ elements, acting on $R=S=\mathbb{C}^n$ via $U_\pi |i\rangle =|\pi(i)\rangle$ for \mbox{$\pi\in S_n$}. Then the measure-and-prepare channel $\mathcal{E}(\sigma_{RS})=\sum_{i=1}^n {\rm tr}(|i\rangle\langle i|_R\otimes\mathbf{1}_S \sigma_{RS})|i_R i_S\rangle\langle i_R i_S|$ is $G$-covariant and satisfies $\mathcal{E}(|j\rangle\langle j|_R\otimes \mathbf{1}_S/n)=|j_R j_S\rangle\langle j_R j_S|$, i.e.\ it maps the $G$-invariant maximally mixed state on $S$ to a $G$-asymmetric pure state on $S$ while leaving the reduced state on $R$ invariant. Intuitively, asymmetry with respect to the permutation group corresponds to classical information which can be cloned and broadcast.

\emph{Weak broadcasting.} While cloning and broadcasting of quantum coherence are impossible, the argument above does not exclude weaker forms of this phenomenon. It is in this setting that we can understand the protocol proposed in Ref.~\cite{aberg2014catalytic}, which we call \emph{weak broadcasting} (it was called `repeatability' in Ref.~\cite{korzekwa2016extraction, cirstoiu2017irreversibility} and `coherence catalysis' in Ref.~\cite{aberg2014catalytic}). The key difference from coherence broadcasting is that the state of the reference $\rho_R$ is allowed to change. Weak broadcasting only requires that the output state in $R$ can be reused in order to induce the same process, arbitrarily many times. More precisely:
\begin{definition}[Weak broadcasting of coherence or asymmetry, Fig.~\ref{fig:weak}]
	\label{def:weakbroad} 
We say that $R$ can weakly broadcast $G$-asymmetry (or \emph{coherence} if $G=\mathbb{R}$) if there exists a system $S$ in a $G$-symmetric state $\rho_S$ and an arbitrary sequence of $G$-covariant operations $(\mathcal{E}^{(n)})_{n\in\mathbb{N}}$ and of states $(\rho_R^{(n)})_{n\in\mathbb{N}}$ such that
\[
   \mathcal{E}^{(n)}\left(\rho_R^{(n)}\otimes\rho_S\right)=\rho_R^{(n+1)}|\rho'_S
\]
for every $n\in\mathbb{N}$, where $\rho'_S$ is $G$-asymmetric.
\end{definition}

In contrast to (strong) broadcasting as introduced in Definition~\ref{DefAsymmetry}, weak broadcasting allows the state of the reference $R$ to change, as long as it does not lose its ability to locally enable the transition $\rho_S\to\rho'_S$ on further initially uncorrelated copies of the system $S$. This is arguably a physically relevant notion, formalizing the most general idea of ``catalytic'' and non-degrading use of a reference frame. Weak broadcasting of coherence is possible --- this is one way to phrase the main result of Ref.~\cite{aberg2014catalytic} (in fact, the explicit protocol presented has $n$-dependence only in $\rho_R$, but not in $\mathcal{E}$). However, the specific scheme proposed exploits an infinite-dimensional source of coherence $\rho_R$ to perform weak broadcasting. A natural question arises: are infinite-dimensional systems truly necessary, or could the same phenomenon arise with finite-dimensional reference frames? In Ref.~\cite[arXiv v1]{cirstoiu2017irreversibility}, it was conjectured that infinite-dimensional coherence sources are necessary. Here we prove that the conjecture holds:
\begin{theorem}
	\label{thm:infinite}If $R$ is finite-dimensional, then weak broadcasting of coherence or $G$-asymmetry is impossible, for any connected Lie group $G$.
\end{theorem}
The proof is given in the Supplemental Material. Its main idea is the observation that, in finite dimensions, the set of covariant channels and the set of states are compact. Thus, if we have weak broadcasting, the sequences $\mathcal{E}^{(n)}$ and $\rho_R^{(n)}$ must have accumulation points, which can be used to construct another state and channel that can be used for broadcasting. That is, in finite dimensions, weak broadcasting implies (strong) broadcasting, which is impossible according to Theorem~\ref{thm:noasymmetrybroadcasting}.

\emph{Application: work extraction from coherence.} Converting coherence among distinct energies of a system $S$ into free energy of an incoherent battery $B$, through energy preserving unitaries on $SB$ and a thermal environment $E$, requires a coherence reference $R$~\cite{Skrzypczyk,lostaglio2015description}. Current proposals ensuring that $R$ does not degrade (in the sense of our definition of weak broadcasting) take $R$ to be a continuous system~\cite{Malabarba}, a doubly-infinite ladder \cite{aberg2014catalytic} or a bounded from below but still infinite ladder~\cite{korzekwa2016extraction}. The question is left open in~\cite{Malabarba} if a finite-dimensional $R$ suffices. Theorem~\ref{thm:infinite} answers it to the negative. 

\emph{Conclusions.} We have shown that broadcasting of coherence or asymmetry is impossible, while weak coherence broadcasting \`a la \r{A}berg necessarily requires infinite-dimensional reference systems -- which is arguably a rather interesting property of a quantum information primitive. Here we have focused on no-go results, but, similarly to the standard no-cloning theorem, one could study in more detail the case in which $R$ is allowed to degrade. Our no-go result does not put any constraints on this scenario, where one expects trade-offs between degradation and creation of asymmetry that depend on the figure of merit of interest.

Our results also find application in the context of ``exotic heat machines'' \cite{Uzdin2018}, specifically those whose aim is to generate energy coherence from thermal resources. By identifying $R$ with the state of the machine, Theorem~\ref{thm:noasymmetrybroadcasting} implies that coherence cannot be created in initially incoherent systems while maintaining the machine (but not necessarily the thermal bath) in a fixed point. This also recovers and generalizes some results of Ref.~\cite{Manzano2018}: in order to amplify coherence in physical systems with a stationary (possibly coherent) machine, these systems must themselves contain some initial coherence. It would be interesting to generalize our results by deriving fundamental bounds on the possibility of amplifying some initial amount of coherence or asymmetry, but such results will have to depend on the choice of coherence measure, in contrast to our fundamental impossibility results.

\emph{Acknowledgments.} We are grateful to Joe Renes for drawing our attention to Ref.~\cite{JanzingBeth}, and to Iman Marivan 
and Rob Spekkens for discussions and for coordinating the submission of their related 
work \cite{MarvianBroad} with us. ML acknowledges financial support from the the European Union's Marie Sklodowska-Curie individual Fellowships (H2020-MSCA-IF-2017, GA794842), Spanish MINECO (Severo Ochoa SEV-2015-0522 and project QIBEQI FIS2016-80773-P), Fundacio Cellex and Generalitat de Catalunya (CERCA Programme and SGR 875). This research was supported in part by Perimeter Institute for Theoretical Physics. Research at Perimeter Institute is supported by the Government of Canada through the Department of Innovation, Science and Economic Development Canada and by the Province of Ontario through the Ministry of Research, Innovation and Science.

\onecolumngrid

\section*{Supplemental Material}

\subsection*{Scope} Our no-go results apply to the `unspeakable' or `energetic' notion of coherence, in which eigenstates like $|0\rangle$ or $|1\rangle$ have physical meaning (e.g.\ as eigenstates of some Hamiltonian or number operator), and where superpositions like $\ket{0} + \ket{1}$ and $\ket{0} + \ket{2}$ are \emph{not} in general equivalent. This should be distinguished \cite{marvian2016how} from a computational (or 'speakable') notion of coherence that has received much attention recently~\cite{streltsov2017colloquium, bu2016catalytic}, in which the labels `$0$', `$1$', `$2$' have no physical significance. As discussed above, it is the processing of unspeakable coherence that is the relevant resource in fields like quantum thermodynamics and quantum metrology. It is within these areas that our results impose fundamental limitations on the processing of superpositions.

\subsection*{Proof of Theorem~1}
Let $[\rho_S,H_S]=0$, and $\rho'_{R S}:=\mathcal{E}(\rho_R\otimes\rho_S)$ with reduced states $\rho'_S$ and $\rho'_R=\rho_R$, and $\mathcal{E}$ a covariant map with respect to $H_S \otimes \mathbf{1}_R + \mathbf{1}_S \otimes H_R$. All systems are taken to be finite-dimensional. We prove the statement by showing that these assumptions imply $[\rho'_S,H_S]=0$.

For $X\in\{S,R,RS\}$ and $t\in\mathbb{R}$, define $\mathcal{U}_t^X(\bullet):=e^{-itH_X}\bullet e^{itH_X}$, and for any state $\sigma_X$ define $\sigma_X(t):=\mathcal{U}_t^X(\sigma_X)$. We have $\mathcal{U}_t^{RS}=\mathcal{U}_t^R\otimes\mathcal{U}_t^S$, and so
\[
   \mathcal{E}(\rho_R(t)\otimes\rho_S) = \mathcal{E}\circ \mathcal{U}_t^{RS}(\rho_R\otimes\rho_S)=\mathcal{U}_t^{R}\otimes\mathcal{U}_t^S \circ \mathcal{E}(\rho_R\otimes\rho_S),
\]
hence ${\rm Tr}_R[\mathcal{E}(\rho_R(t)\otimes\rho_S)]=\rho'_S(t)$ and ${\rm Tr}_S[\mathcal{E}(\rho_R(t)\otimes\rho_S)]=\rho_R(t)$. It follows that if we define
\[
   \mathcal{T}(\sigma_R):={\rm Tr}_S[\mathcal{E}(\sigma_R\otimes\rho_S)]
\]
we have $\mathcal{T}(\rho_R(t))=\rho_R(t)$ for all $t\in\mathbb{R}$. 

From the above, coherence broadcasting implies the existence of a quantum operation $\mathcal{E}$ and a state $\rho_S$ with $\rho_S(t)$ constant in $t$ satisfying (using the notation introduced in the main text)
\begin{equation}
\label{eq:reformulation}
\mathcal{E}(\rho_R(t) \otimes \rho_S) = \rho_R (t)|\rho'_S(t),
\end{equation}
with $\rho'_S(t)$ not constant in $t$. 

Let us then focus on Eq.~\eqref{eq:reformulation}. Let $\tilde R$ be the smallest subspace of $R$ that supports all the $\rho_R(t)$, i.e.\ $\tilde R:={\rm span}\bigcup_{t\in\mathbb{R}}{\rm supp}\,\rho_R(t)$. There is a finite subset $T\subset\mathbb{R}$ such that the state $\bar\rho_R:=\frac 1 {|T|}\sum_{t\in T}\rho_R(t)$ has ${\rm supp}\,\bar\rho_R=\tilde R$; moreover, $\mathcal{T}(\bar\rho_R)=\bar\rho_R$. Denoting the orthogonal projector onto $\tilde R$ by $\Pi_{\tilde R}$, it follows from~\cite[Prop.\ 6.10]{Wolf} that $\sigma_R\leq\Pi_{\tilde R}$ implies $\mathcal{T}(\sigma_R)\leq \Pi_{\tilde R}$. That is, we can consider $\mathcal{T}$ as a channel on the states of $\tilde R$ by restricting its domain of definition to this subspace. Let us define $R'$ to be the orthogonal complement of $\tilde R$ in $R$, i.e.\ $R=\tilde R \oplus R'$. Furthermore, for any subspace $X$, let $\Pi_X$ denote the orthogonal projector onto $X$. Define a new channel $\mathcal{\tilde E}$ on $\tilde R S \equiv \tilde R \otimes S \subseteq R\otimes S$ via
\[
   \mathcal{\tilde E}(\rho_{\tilde R S}):=\Pi_{\tilde R S}\mathcal{E}(\rho_{\tilde R S})\Pi_{\tilde R S}+\tr{}{\Pi_{R'S}\mathcal{E}(\rho_{\tilde R S})}\cdot \frac{\mathbf{1}_{\tilde R S}}{d_{\tilde R S}},
\]
where $d_{\tilde R S}=\dim(\tilde R \otimes S)$. This is a completely positive trace-preserving map. To compute its action on $\rho_R(t)\otimes\rho_S$, note that
\[
   \tr{}{\Pi_{\tilde R S}\rho_R(t)|\rho'_S(t)}=\tr{}{\Pi_{\tilde R}\otimes\mathbf{1}_S \rho_R(t)|\rho'_S(t)}=\tr{R}{\Pi_{\tilde R} {\rm Tr}_S\left(\rho_R(t)|\rho'_S(t)\right)}=\tr{R}{\Pi_{\tilde R}\rho_R(t)}=1.
\]
Thus ${\rm Tr}\left[\Pi_{R' S}\mathcal{E}(\rho_R(t)\otimes\rho_S)\right]=0$ and $\Pi_{\tilde R S}\mathcal{E}(\rho_R(t)\otimes\rho_S)\Pi_{\tilde R S}=\rho_R(t)|\rho'_S(t)$. This shows that
\[
   \mathcal{\tilde E}(\rho_R(t) \otimes \rho_S) = \rho_R (t)|\rho'_S(t).
\]
In other words, we can replace $R$ by $\tilde R$ and $\mathcal{E}$ by $\mathcal{\tilde E}$, and we still have Eq.~\eqref{eq:reformulation}. Moreover, the minimal subspace supporting all $\rho_R(t)$ is $\tilde R$, and $\mathcal{\tilde T}(\sigma_{\tilde R}):={\rm Tr}_S[\mathcal{\tilde E}(\sigma_{\tilde R}\otimes\rho_S)]$ defines a channel on $\tilde R$, since it is the restriction of $\mathcal{T}$ to $\tilde{R}$. Clearly, $\mathcal{\tilde T}(\rho_R(t))=\rho_R(t)$ for all $t \in \mathbb{R}$. This allows us to use the results of~\cite{Hayden2004}; we will drop the tildes on $\tilde R$, $\mathcal{\tilde E}$ and $\mathcal{\tilde T}$ from now on. Consider any (not necessarily covariant) Stinespring dilation of the channel $\mathcal{E}$, i.e.\ an ancillary system $E$, a unitary $U_{RAE}$ and a state $\rho_E$ such that
\[
   \mathcal{E}(\sigma_{RS})={\rm Tr}_E\left[ U_{RSE}(\sigma_{RS}\otimes\rho_E)U_{RSE}^\dagger\right].
\]
Now we use the Koashi-Imoto decomposition~\cite{KoashiImoto} as presented in~\cite{Hayden2004}. Since $\mathcal{T}(\rho_R(t))=\rho_R(t)$ for every $t$, it says that there exists a decomposition of the form
\[
   R = \bigoplus_j \mathcal{J}_j\otimes\mathcal{K}_j,\qquad \rho_R(t)=\bigoplus_j q_{j|t} \rho_{j|t}\otimes \omega_j,\qquad
   U_{RSE}=\bigoplus_j \mathbf{1}_{\mathcal{J}_j}\otimes V_{\mathcal{K}_j SE},
\]
where the $(q_{j|t})_j$ are probability distributions over $j$, every $\rho_{j|t}$ is a state on $\mathcal{J}_j$, $\omega_j$ is a state on $\mathcal{K}_j$ (independent of $t$). For every $t\in\mathbb{R}$ and every $j$, define ${\rm spec}\, \rho_R^{(j)}(t)$ as the vector of eigenvalues of $\rho_R^{(j)}(t):=q_{j|t} \rho_{j|t}\otimes\omega_j=\Pi_{\mathcal{J}_j\otimes\mathcal{K}_j}\rho_R(t)\Pi_{\mathcal{J}_j\otimes\mathcal{K}_j}$ in non-increasing order. According to Weyl's Perturbation Theorem~\cite{Bhatia}, this expression is continuous in $t$. Since the spectrum of $\rho_R(t)$ is independent of $t$, the entries of ${\rm spec}\, \rho_R^{(j)}(t)$ are a subset of the discrete set of eigenvalues of $\rho_R=\rho_R(0)$. Consequently, continuity forces ${\rm spec}\, \rho_R^{(j)}(t)$ to be constant in $t$. Hence, $q_{j|t}={\rm tr}\, \rho_R^{(j)}(t)$ is independent of $t$, and will henceforth be called $q_j$. We obtain
\begin{eqnarray*}
\rho'_S(t)&=& {\rm Tr}_{RE}\left[U_{RSE}\left(\rho_R(t)\otimes\rho_S\otimes\rho_E\right)U_{RSE}^\dagger\right]
={\rm Tr}_{RE} \left[ \bigoplus_j q_j \rho_{j|t} \otimes V_{\mathcal{K}_j S E}\left(\omega_j\otimes\rho_S\otimes\rho_E\right)V_{\mathcal{K}_j S E}^\dagger\right]\\
&=& \sum_j q_j {\rm Tr}_{\mathcal{K}_j E}\left[V_{\mathcal{K}_j S E} \left(\omega_j\otimes\rho_S\otimes\rho_E\right)V_{\mathcal{K}_j S E}^\dagger\right],
\end{eqnarray*}
where in the second line we used the relation ($R_j \equiv \mathcal{J}_j \otimes \mathcal{K}_j$)
\[
{\rm Tr}_{\oplus_j R_j E} \left(\oplus_j Q_{R_j S E}  \right) = \sum_j {\rm Tr}_{R_j E} \left(Q_{R_j S E}\right),
\]
where each $Q_{R_j SE}$ is an operator on $R_jSE$. This expression is independent of $t$, hence $\rho'_S(t)=\rho'_S(0)$ for all $t$, which implies that $[\rho'_S,H_S]=0$.
\qed

\subsection*{Coherence broadcasting vs.\ broadcasting} If $\mathcal{E}$ is any quantum operation (not necessarily covariant) and Eq.~\eqref{eq:reformulation} holds (for all $t$), then one can check that the definition of coherence broadcasting is satisfied for the covariant map $\mathcal{\tilde E}:=\lim_{T\to\infty} \frac 1 T \int_0^T \mathcal{U}_{-t}\circ\mathcal{E}\circ\mathcal{U}_t\, {\rm d}t$ (diagonalization and direct integration shows that this limit exists, and that this map is covariant), i.e.\ $\mathcal{\tilde E}(\rho_R\otimes\rho_S)=\rho_R|\rho'_S$. Conversely, as we proved above, Eq.~\eqref{eq:reformulation} follows from the definition of coherence broadcasting. We conclude that Eq.~\eqref{eq:reformulation} is \emph{equivalent} to coherence broadcasting. Note in passing that, if we fixed $S \equiv R$ and required $\rho'_S(t) \equiv \rho_R(t)$, then coherence broadcasting would be equivalent to (standard) broadcasting of the states $\{\rho_R(t)\}_{t \in \mathbb{R}}$ (a minor remark: if the energies in $H_R$ are rational and $R$ is finite-dimensional, then $t$ will range only within a finite interval). Due to the extra freedom, however, we cannot prove the no coherence broadcasting theorem from the standard no broadcasting theorem.

\subsection*{Cloning timing (and asymmetry) information in the classical limit} If we consider the coherent state $\ket{\alpha}_R$, then $\{\ket{\alpha(t)}_R\}_{t \in \mathbb{R}}$ become all mutually orthogonal in the limit $|\alpha| \rightarrow \infty$. From the equivalence of coherence cloning with Eq.~\eqref{eq:reformulation}, and the fact that mutually orthogonal states can be cloned, it follows that as we approach the classical limit we can clone the timing information arbitrarily well, i.e.\ coherence can be cloned with arbitrary precision (see also \cite{cirstoiu2017irreversibility}). The same holds true for every sequence of $\rho^{(n)}_R$ such that $\rho^{(n)}_R(t)$ become arbitrarily close to a family of mutually orthogonal states as $n \rightarrow \infty$. Our no-go theorem is then compatible with the \emph{classical} intuition that timing information (as well as other asymmetry information, such as directional information) can be distributed without degrading the corresponding reference frame.

\subsection*{Proof of Theorem~2} The proof is analogous to the proof of Theorem~1 with the following modifications. First, replace every $\mathcal{U}_t^X$ and $\sigma_X(t)$ by $\mathcal{U}_g^X$ and $\sigma_X(g)$, with $g \in G$, and replace the equations $[\rho_A,H_A]=0$ and $[\rho'_A,H_A]=0$ by the statements that $\rho_A$ resp.\ $\rho'_A$ are $G$-invariant. Furthermore, $q_{j|t}$ will be replaced by $q_{j|g}$, and $\rho_{j|t}$ by $\rho_{j|g}$. Note that the assumption of connectedness of $G$ becomes relevant in the argument which uses continuity of ${\rm spec}\, \rho_B^{(j)}(g)$ to conclude that these spectra are constant in~$g$.
\qed

\subsection*{Further comments on Theorem~2} At first glance it may seem as if there was a mistake in the formulation of the theorem: the trivial group $G=\{\mathbf{1}\}$ is a connected Lie group; but isn't $G$-asymmetry broadcasting possible for this group $G$? The answer is: no, it is not. That is because \emph{all} states are $G$-symmetric under the trivial group $G$, hence no operation can create any $G$-asymmetry, and $G$-asymmetry broadcasting is trivially impossible.

There is another question that suggests itself at first glance: should we really demand in the formulation of the theorem that $G$ is a connected Lie group? Wouldn't it be sufficient to demand that $G$ has a non-trivial connected component at the identity (call it $G_0$)? For example, we could have $G={\rm O}(3)$ and $G_0={\rm SO}(3)$. Since $G_0$-asymmetry broadcasting is impossible, doesn't this imply that $G$-asymmetry broadcasting is impossible?

Unfortunately, this implication is not in all cases correct. To see this, consider the example $G={\rm U}(1)\times S_n$, where $S_n$ is the permutation group of $n$ elements. In this case, $G_0=\{(e^{i\theta},\mathbf{1})\}$ is non-trivial, but consider the group representation $U_g=\pi$ for $g=(e^{i\theta},\pi)$, representing $G$ on $\mathbb{C}^n$. Using two copies of this representation, the discrete group asymmetry broadcasting example from the main text shows that $G$-asymmetry broadcasting is possible. This is clearly a very special case, owing to the fact that $G$ contains the permutation group as a normal subgroup. Nevertheless, it disproves the conjecture that we have just described. Since connected Lie groups are the physically most interesting groups, we do not pursue this analysis (of more general Lie groups) further at this point.

\subsection*{Proof of Theorem~3}
Assume weak broadcasting as in the statement of the theorem. Since the set of density matrices on $R$ is compact, there exists a convergent subsequence $(\rho_R^{(n_k)})_{k\in\mathbb{N}}$. Set $\sigma_R^{(k)}:=\rho_R^{(n_k)}$ and $\sigma_R:=\lim_{k\to\infty}\sigma_R^{(k)}$. For every $m\in\mathbb{N}$, we now define a channel $\mathcal{C}_m$ from states on $RS$ to states on $RS$ as follows. Consider many copies of the system $S$, labeled by $S_{n_m}, S_{n_m+1},\ldots, S_{n_{m+1}-1}$ (these are $n_{m+1}-n_m$ many). We also introduce the special label $S:=S_{n_m}$. Now we act with a suitable sequence of $\mathcal{E}$-channels on $R S_i$ for $i=n_m, \ldots n_{m+1}-1$; every use of the suitable $\mathcal{E}$ preserves the usability of the state on $R$ and transforms the local state in $S_i$ into $\rho'_S$. Formally, we define $\mathcal{C}_m$ as follows, for any state $\tau_{RS}$ on $RS$:
\[
   \mathcal{C}_m(\tau_{RS}):={\rm Tr}_{S_{n_m} S_{n_m+1}\ldots S_{n_{m+1}-2}}\left[
      \mathcal{E}^{(n_{m+1}-1)}_{R S_{n_{m+1}-1}}\circ \ldots\circ \mathcal{E}_{R S_{n_m+1}}^{(n_m+1)}\circ \mathcal{E}_{R S_{n_m}}^{(n_m)}\left(
         \tau_{RS}\otimes\rho_{S_{n_m+1}}\otimes \rho_{S_{n_m+2}}\otimes\ldots\otimes\rho_{S_{n_{m+1}-1}}
      \right)
   \right],
\]
where the subscript at the $\mathcal{E}$ denotes the tensor factors on which they act, and the $\rho_{S_i}$ are copies of a fixed $\rho_S$ with $[\rho_S,H_S] = 0$ at the corresponding tensor factors. For clarity, here is an example: if e.g.\ $n_1=2$ and $n_2=5$, then
\[
   \mathcal{C}_1(\tau_{RS})\equiv \mathcal{C}_1(\tau_{R S_2})=
   {\rm Tr}_{S_2 S_3}\left[
      \mathcal{E}^{(4)}_{R S_4}\circ \mathcal{E}^{(3)}_{R S_3}\circ \mathcal{E}^{(2)}_{R S_2}\left(
         \tau_{R S_2}\otimes\rho_{S_3}\otimes\rho_{S_4}
      \right)
   \right].
\]
In this example, by first performing the partial trace over $S_2$, we obtain
\[
   \mathcal{C}_1\left(\rho_R^{(2)}\otimes \rho_{S_2}\right)={\rm Tr}_{S_3}\left[
      \mathcal{E}_{R S_4}^{(4)}\circ \mathcal{E}_{R S_3}^{(3)}\left(\rho_R^{(3)}\otimes\rho_{S_3}\otimes \rho_{S_4}\right)
   \right]
      =\mathcal{E}_{R S_4}^{(4)}\left(\rho_R^{(4)}\otimes \rho_{S_4}\right)=\rho_R^{(5)}|\rho'_S.
\]
Generalizing this example, we have
\[
   \mathcal{C}_m\left(\sigma_R^{(m)}\otimes\rho_S\right)=\mathcal{C}_m\left(\rho_R^{(n_m)}\otimes\rho_{S_{n_m}}\right)=\rho_R^{(n_{m+1})}|\rho'_S=\sigma_R^{(m+1)}|\rho'_S.
\]
Moreover, every channel $\mathcal{C}_m$ is $G$-covariant, since it is a composition of covariant channels and the ancillas used are symmetric. Taking again the previous example for the sake of clarity, we have for every $g \in G$
	   \begin{align*} 
	   \mathcal{C}_1(\mathcal{U}^{R}_g \otimes \mathcal{U}^{S_2}_g (\tau_{R S_2}))&= {\rm Tr}_{S_2 S_3}\left[
	   \mathcal{E}^{(4)}_{R S_4}\circ \mathcal{E}^{(3)}_{R S_3}\circ \mathcal{E}^{(2)}_{R S_2} \mathcal{U}^{R}_g \otimes \mathcal{U}^{S_2}_g \otimes \mathcal{U}^{S_3}_g \otimes \mathcal{U}^{S_4}_g\left(
	   \tau_{R S_2}\otimes\rho_{S_3}\otimes\rho_{S_4}
	   \right)
	   \right]
	   \\ & =
	    {\rm Tr}_{S_2 S_3}\left[\mathcal{U}^{R}_g \otimes \mathcal{U}^{S_2}_g \otimes \mathcal{U}^{S_3}_g \otimes \mathcal{U}^{S_4}_g
	   \mathcal{E}^{(4)}_{R S_4}\circ \mathcal{E}^{(3)}_{R S_3}\circ \mathcal{E}^{(2)}_{R S_2} \left(
	   \tau_{R S_2}\otimes\rho_{S_3}\otimes\rho_{S_4}
	   \right)
	   \right]\\ & = \mathcal{U}^{R}_g \otimes \mathcal{U}^{S_4}_g \mathcal{C}_1(\tau_{R S_2}),
	\end{align*}  
	which, by identification of the spaces $S_i$, can also be rewritten as $\mathcal{C}_1(\mathcal{U}^{R}_g \otimes \mathcal{U}^{S}_g (\tau_{R S})) = \mathcal{U}^{R}_g \otimes \mathcal{U}^{S}_g \mathcal{C}_1(\tau_{R S})$. The same holds for every $\mathcal{C}_m$.
	 
The set of $G$-covariant channels on the states of a finite-dimensional Hilbert space is topologically closed. To show this, let $\|X\|:={\rm tr}|X|$ be the trace norm, and $\mathcal{M}_n$ a sequence of covariant channels converging to $\mathcal{M}$. Without loss of generality, we consider the induced norm $\|\mathcal{M}\|:=\sup_{\|X\|=1} \|\mathcal{M}(X)\|$. Then, for every $\rho$ and every $g \in G$,
\begin{align*}
\| \mathcal{U}_g(\mathcal{M}(\rho)) -\mathcal{M}(\mathcal{U}_g(\rho))\| &\leq  \| \mathcal{U}_g(\mathcal{M}(\rho)) -\mathcal{M}_n(\mathcal{U}_g(\rho))\| + \| \mathcal{M}_n(\mathcal{U}_g(\rho)) -\mathcal{M}(\mathcal{U}_g(\rho))\|  \\
& \leq \|\mathcal{U}_g(\mathcal{M}(\rho))-\mathcal{U}_g(\mathcal{M}_n(\rho))\|+\|\mathcal{M}_n-\mathcal{M}\|=\|\mathcal{M}(\rho)-\mathcal{M}_n(\rho)\|+\|\mathcal{M}-\mathcal{M}_n\|\\
&\leq 2 \|\mathcal{M}-\mathcal{M}_n\|
\stackrel{n \rightarrow \infty}{\longrightarrow} 0,
\end{align*}
where for the first inequality we used the triangle inequality, for the second inequality we used the covariance of $\mathcal{M}_n$ and the definition of the induced norm, for the third inequality we used the unitary invariance of the trace norm, and for the limit we used the assumption that $\mathcal{M}_n$ converges to $\mathcal{M}$. This proves that $\mathcal{U}_g(\mathcal{M}(\rho))=\mathcal{M}(\mathcal{U}_g(\rho))$, i.e.\ that the limit channel is covariant. Finally, since the full set of channels is compact (and thus bounded), so is the set of $G$-covariant channels.

Since the set of covariant channels is compact, the sequence $(\mathcal{C}_m)_{m\in\mathbb{N}}$ has a convergent subsequence $(\mathcal{C}_{m_k})_{k\in\mathbb{N}}$. Define $\mathcal{C}:=\lim_{k\to\infty} \mathcal{C}_{m_k}$, then $\mathcal{C}$ is covariant and satisfies $\mathcal{C}(\sigma_R\otimes\rho_S)=\sigma_R|\rho'_S$ (since $\sigma^{(k)}_R$ converges to $\sigma_R$). That is, we have strong broadcasting of $G$-asymmetry, which is impossible according to Theorem~2.
\qed

\subsection*{Comment on infinite-dimensional reference systems $R$} While weak broadcasting is impossible for finite-dimensional references according to Theorem~3, \r{A}berg's result~\cite{aberg2014catalytic} shows that it can be accomplished with infinite-dimensional $R$. At first sight, there is a natural intuition for why the change of state of $R$ seems crucial to achieve broadcasting, which is reminiscent of the idea of ``Hilbert's hotel'': similarly as we can always create a free room in Hilbert's hotel by shifting all existing guests towards infinity, we can think of extracting ever more coherence (or, more generally, $G$-asymmetry) from $R$ by pushing its quantum state towards energetic infinity (increasing its support indefinitely). This operation, as the argumentation goes, would basically ``steal'' some of the $G$-asymmetry of $R$ (of which there is, in some sense, an infinite amount) and move it to the system of interest. This seems impossible without changing the state of $R$. Therefore, this intuition would suggest that \emph{strong} broadcasting of coherence or asymmetry (in the sense of Definition~2) will most likely be impossible.

However, at closer inspection, this intuitive argumentation does not fully hold up. First, while Hilbert's hotel starts out with a literally infinite supply of resources (i.e.\ empty rooms), \r{A}berg's protocol begins in a well-defined quantum state with finite support that does not in any obvious way hold an infinite amount of coherence. Second, there are results in the literature which show that infinite-dimensional reference systems \emph{can} act as ``strong'' catalysts along the lines we demand (though in a different context that does not involve asymmetry). Namely, in Ref.~\cite{cleve2017perfect}, it is shown that infinite-dimensional $R$ allow for ``perfect embezzlement'' of quantum entanglement. That is, consider two agents $A$ and $B$ that start with local states $|0\rangle_A$ and $|0\rangle_B$ that they want to transform into an entangled state. Furthermore, suppose that they share a reference system $R$ and that $A$ ($B$) can perform a set of operations that commutes with those of $B$ ($A$) (this restricts them to ``local operations'' in a specific sense). Then, they can implement the transformation
\[
   |0\rangle_A\otimes |0\rangle_B\otimes|\psi\rangle_R \mapsto |\varphi_+\rangle_{AB}\otimes |\psi\rangle_R
\]
exactly, where $|\varphi_+\rangle_{AB}$ is a maximally entangled state on $AB$, and $|\psi\rangle_R$ is an entangled state of the reference that acts as a catalyst. (Note that this is different from the well-known results on finite-dimensional embezzling by van Dam and Hayden~\cite{vanDamHayden}, in which the state of the catalyst is only approximately preserved). Applying our terminology to this setup, we could say that this is a case where the resource of entanglement is strongly broadcast (in fact, cloned). This example suggests that strong broadcasting of asymmetry, even though intuitively unlikely, may not be impossible for infinite-dimensional references in all cases.

This example also shows that the behavior of the infinite-dimensional case may depend strongly on its detailed mathematical formulation, reflecting alternative possibilities to model the physical control of the quantum systems. Namely, it is shown in Ref.~\cite{cleve2017perfect} that this sort of perfect embezzling is only possible if the notion of locality on $R$ is defined in terms of commuting operators (resembling the usual approach of quantum field theory), and \emph{not} in terms of tensor products of Hilbert spaces. This warns us that the (im)possibility of broadcasting of $G$-asymmetry may well depend on the details of the mathematical formulation, like the types of operator algebras we allow for $R$ and assumptions of how the group $G$ is allowed to act on it.

\subsection*{Revised conjecture on correlating thermal machines} As mentioned in the main text, the following result is proven in Ref.~\cite{Mueller2017}: \emph{if $\rho_A$ and $\rho'_A$ are block-diagonal quantum states, then for every $\epsilon>0$ there exists a finite-dimensional catalyst $\sigma_R$, a thermal operation $\mathcal{T}$, and a state $\rho'_A(\epsilon)$ with $\|\rho'_A(\epsilon)-\rho'_A\|<\epsilon$ such that}
\begin{equation}
   \mathcal{T}(\rho_A\otimes\sigma_R)= \rho'_A(\epsilon)|\sigma_R
   \label{eqTau}
\end{equation}
\emph{if and only if $F(\rho_A)\geq F(\rho'_A)$.} It has been conjectured in Ref.~\cite{Mueller2017} that this result is also true for states that are not block-diagonal, i.e.\ in the presence of coherence.

However, Theorem~1 disproves this conjecture, which can be seen as follows. Choose any pair of states $\rho_A,\rho'_A$ with $F(\rho_A)\geq F(\rho'_A)$ such that $[\rho_A,H_A]=0$ (i.e.\ $\rho_A$ is block-diagonal) but $[\rho'_A,H_A]\neq 0$ (and hence $[\rho'_A(\epsilon),H_A]\neq 0$ if $\epsilon$ is small enough). Then Eq.~\eqref{eqTau} would be a case of coherence broadcasting which is impossible.

This argument rests crucially on the fact that every $\mathcal{T}$ is a thermal operation and thus covariant. However, there is another set of operations of thermodynamic relevance which enables the same set of transitions between pairs of \emph{block-diagonal} states, but admits more transitions between states with coherence~\cite{faist2015gibbs}: these are the \emph{Gibbs-preserving maps}, i.e.\ the completely positive, trace-preserving maps $\mathcal{T}$ with $\mathcal{T}(\gamma_X)=\gamma_X$ for the thermal state $\gamma_X$. Not all Gibbs-preserving maps are covariant, hence the argument above does not apply to them. This suggests the following revised conjecture:
\begin{conjecture}[Free energy for states with coherence] Consider a pair of quantum states $\rho_A,\rho'_A$ on a finite-dimensional quantum system $A$. Then, for every $\epsilon>0$ there exists a finite-dimensional catalyst $\sigma_R$, a Gibbs-preserving operation $\mathcal{G}$, and a state $\rho'_A(\epsilon)$ with $\|\rho'_A(\epsilon)-\rho'_A\|<\epsilon$ such that
\[
   \mathcal{G}(\rho_A\otimes\sigma_R)= \rho'_A(\epsilon)|\sigma_R
\]
if and only if $F(\rho_A)\geq F(\rho'_A)$.
\end{conjecture}

\subsection*{Relation to the ``No-Local-Broadcasting Theorem for Multipartite Quantum Correlations'' by Piani et al.~\cite{Piani}}
In Ref.~\cite{Piani}, the authors present a different kind of no-broadcasting result which can be related to some special cases of our work. To understand this relation, recall that the asymmetry of a quantum reference frame can be treated externally (i.e., be defined with respect to an implicit classical reference, as we did here) or explicitly (through correlations with the quantum state of this classical reference). It is shown in Ref.~\cite{bartlett2007reference} that both pictures are equivalent. We will now discuss the case of a finite group $G$, but we conjecture that similar conclusions can be drawn for connected Lie groups by a suitable limit procedure.

Treating the reference $R$ explicitly, we model it by saying that it is in some state $\rho_R(g)$, where $g\in G$ is a random variable carried by a classical reference frame $C$, described by a $|G|$-dimensional Hilbert space with orthonormal basis $\{|g\rangle\}_{g\in G}$. That is,
\[
   \rho_{CR}=\frac 1 {|G|}\sum_{g\in G} |g\rangle\langle g|_C\otimes \rho_R(g).
\]
Note that tracing out $C$ produces a $G$-symmetric state on $R$.

Suppose that we have an additional system $S$ in a $G$-symmetric state $\rho_S$. As we have shown in the main text, if $G$-asymmetry broadcasting is possible, then there exists a quantum operation that transforms $\rho_R(g)\otimes \rho_S$ into $\rho_R(g)|\rho'_S(g)$, for all $g\in G$. Applying this map to one half of the state $\rho_{CR}$, and at the same time copying the classical register $C$ into another classical register $C'$, gives us a transition
\[
   \frac 1 {|G|} \sum_{g\in G}|g\rangle\langle g|_C\otimes \rho_R(g)\otimes \rho_S \mapsto  \frac 1 {|G|} \sum_{g\in G} |g\rangle\langle g|_C\otimes |g\rangle\langle g|_{C'}\otimes\rho_R(g)|\rho'_S(g)=:\rho_{CC'RS}
\]
via some quantum operation.

Consider the special case that $S$ has the same Hilbert space dimension as $R$, and that $\rho'_S=\rho_R(\mathbf{1})$, where $\mathbf{1}\in G$ is the unit element of $G$. If this is the case, then $\rho_{CC'RS}$ is a \emph{broadcast state} for $\rho:=\frac 1 {|G|} \sum_{g\in G}|g\rangle\langle g|\otimes \rho_R(g)$ in the sense of Ref.~\cite{Piani}, since $\rho_{CR}=\rho_{C'S}=\rho$. The existence of the aforementioned quantum operation thus implies that $\rho$ is \emph{locally broadcastable}. But then, Theorem 3 in Ref.~\cite{Piani} shows that $\rho$ is a ``classical-classical'' state, i.e.
\begin{equation}
   [\rho_R(g),\rho_R(g')]=0\quad\mbox{if }g\neq g'.
   \label{eqCommute}
\end{equation}
For finite groups $G$, this is possible if $\rho_R(g)=|g\rangle\langle g|$, i.e.\ if $R$ is also a classical reference. Assuming that the argumentation above can be generalized to connected non-trivial Lie groups $G$, eq.~(\ref{eqCommute}) tells us that, for such groups, the reference $R$ has to be infinite-dimensional. This establishes no broadcasting of asymmetry (our Theorem 2), however \emph{only in the special case} that $\rho'_S=\rho_R(\mathbf{1})$, i.e.\ $\rho'_S=\rho_R$ in our definition of asymmetry broadcasting (Definition 2). However, this special case can be more directly excluded via the traditional no-broadcasting theorem.

In fact, using Theorem 3 of Ref.~\cite{Piani}, one can obtain a slightly stronger result which is, however, still weaker than our Theorem 2. This result of Ref.~\cite{Piani} says that the transition above is impossible if we demand $I(C:R)=I(C':S)$ for the mutual information (which is weaker than $\rho_{CR}=\rho_{C'S}$). A simple calculation shows that $I(C:R)=A(\rho_R(\mathbf{1}))$, where $A$ is the ``relative entropy of frameness'', a measure of asymmetry defined in Ref.~\cite{Gour} as
\begin{equation}
A(\sigma) = S(\sigma \| \mathcal{G}(\sigma)), \quad \mathcal{G}(\sigma) = \int dg U_g \sigma U^\dag_g.
\end{equation}
 Repeating the argumentation from above, this tells us that $G$-asymmetry broadcasting \emph{under the additional constraint} that $A(\rho_R)=A(\rho'_S)$ is impossible.

\end{document}